\documentclass[prl,twocolumn,superscriptaddress,floatfix]{revtex4-1}
\usepackage{epsfig}
\usepackage{latexsym}
\usepackage{amsmath}
\usepackage{graphics}
\usepackage{amsfonts}
\usepackage{float}

\usepackage{color}
\usepackage{pstricks}

\newcommand{\br}{\mathbf{r}}

\begin{document}

\title{Casimir-Polder interaction between a polarizable particle and a plate with a hole}
\author{Claudia Eberlein}
\author{Robert Zietal}
\affiliation{Department of Physics \& Astronomy,
    University of Sussex,
     Falmer, Brighton BN1 9QH, England}
\date{\today}
\begin{abstract}
We determine exactly the non-retarded Casimir-Polder interaction between a neutral but polarizable particle and a perfectly reflecting sheet containing a circular hole. The calculation reveals a strong dependence of the interaction on the orientation of the particle's electric dipole moment with respect to the surface. For a dipole moment that is polarized perpendicularly to the surface the interaction potential has two saddle points lying above and below the plane of the surface, on a line through the centre of the aperture. For a dipole moment that is polarized parallel to the surface there is only one saddle point right in the middle of the aperture. Provided the particles' motion could be confined to a line through the middle of the aperture, this effect could potentially be used for population-sensitive trapping of particles.
\end{abstract}


\maketitle
\begin{figure}[b]
\includegraphics[width=6.0 cm, height=3.5 cm]{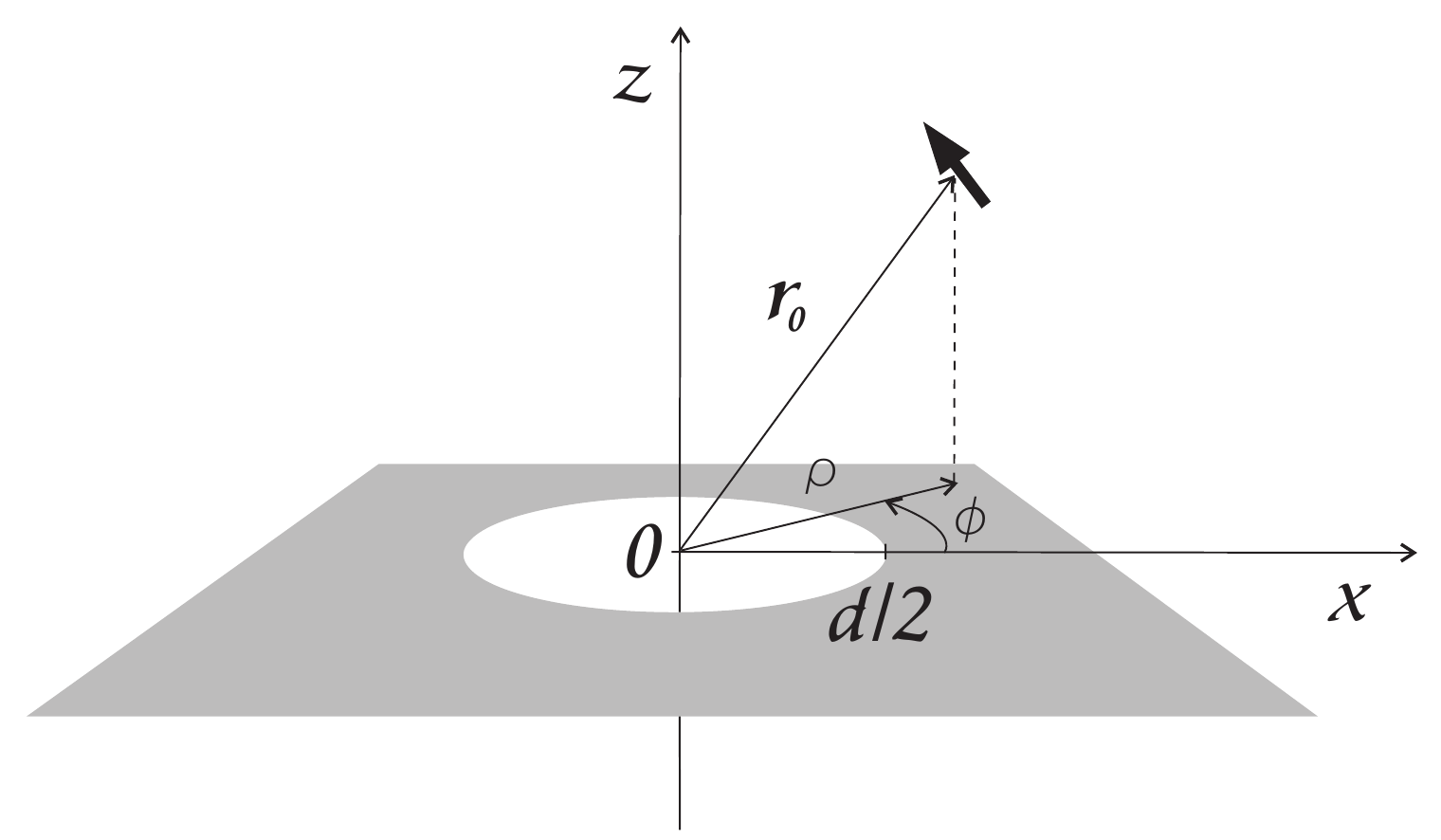}
\caption{\label{fig:geometry} (Color online) A dipole at a position $\br_0$ in the vicinity of the perfectly reflecting sheet with a circular aperture of diameter $d$.}
\end{figure}

It is well-known that the interaction of an atom with a surface depends on the orientation of the atomic dipole moment with respect to the surface. For a ground-state atom and a flat, perfectly-reflecting surface this difference manifests itself only in marginally different pre-factors in the same kind of analytical expression \cite{CP}. However, for more complicated structures such effects may be more pronounced. Recently we have shown that the large-distance behaviour of the interaction between an atom a cylindrical wire may display either logarithmic or power-like behaviour depending on how the atomic dipole is oriented with respect to the wire \cite{Micro1,Micro2}. Furthermore, it has also turned out that there is no Casimir-Polder attraction for an atom placed exactly above the edge of a semi-infinite sheet of a conductor provided the atomic dipole moment is perpendicular to the material's surface \cite{Micro1,Micro2}. The question arises as to whether this conclusion would continue to hold if we were to add to the system some other (infinitely) thin, flat conductors lying in the same plane. In particular, would the energy shift of a $z$-polarized dipole vanish when the dipole is located in a circular aperture of a perfectly reflecting sheet? (cf. Fig. \ref{fig:geometry}). 
This is interesting because we know that the {atom-plate} interaction energy at very large distances, where the presence of the aperture can be neglected, approaches zero from below \cite{CP}. Thus, the vanishing of the energy shift at $z=0$ within the aperture would imply that the potential calculated along any path that is perpendicular to the plate and falls within the aperture features a minimum at some point \cite{Levin}. Such a minimum would imply some intricate behaviour of the Casimir-Polder force acting on a $z$-polarized dipole close to plate with a hole. Therefore, it seems worthwhile to investigate the problem in detail. To capture the essential features of the problem we consider the non-retarded limit of the Casimir-Polder interaction, that is, we assume that the atom-surface separation is much smaller than the wavelength of the strongest dipole transition within the atom. This is the regime of the greatest practical interest, especially in the context of cold-atom experiments whose aim is to trap and guide atoms very close to surfaces. The non-retarded limit of the Casimir-Polder interaction between a neutral atom and a perfectly reflecting microstructure can be worked out by methods of electrostatics. Envisaging the atom to be a point-like electric dipole located at $\br_0$ one can show that the interaction energy with a nearby surface can be written in rectangular coordinates as \cite{Micro1}
\begin{equation}
\Delta E = \frac{1}{2\epsilon_0}\lim_{\br,\br' \rightarrow\br_0} \sum_{i=1}^3\langle \mu_i^2 \rangle\nabla_i\nabla'_i G_H(\br,\br')\label{eqn:EnergyShift}
\end{equation}
where $G_H(\br,\br')$ is the homogeneous part of the electrostatic potential at $\br$ due to a point charge at $\br'$. The sum runs over the three components of the electric dipole moment operator $\mu_i$. The potential $G(\br,\br')$ due to a unit point charge at $\br'$ satisfies
\begin{equation}
-\nabla^2G(\br,\br')=\delta^{(3)}(\br-\br')\label{eqn:Poisson}
\end{equation}
with appropriate boundary conditions. In the case of a perfect reflector, $G(\br,\br')$ would have to vanish for any $\br$ on its surface. The homogeneous part of the potential is the difference between $G(\br,\br')$ and the potential of the same point charge in free space, i.e. without any boundary conditions,
\begin{equation}
G_H(\br,\br')=G(\br,\br')-\frac{1}{4\pi}\frac{1}{|\br-\br'|}.
\end{equation}
So far the problem is entirely classical and does not involve quantum electrodynamics. The quantum properties of an atom are accounted for in the expectation values of the electric dipole moment operator $\langle\mu^2_i\rangle\equiv\langle j |\mu_i^2|j\rangle$, where $|j\rangle$ denotes the atomic state, not necessarily its ground state. The difficulty of determining the energy shift (\ref{eqn:EnergyShift}) lies in calculating the potential $G(\br,\br')$ of a point charge, also called a Green's function, for the geometry of interest. In our case we need the solution of (\ref{eqn:Poisson}) for point charge placed close to a plate with a circular hole of diameter $d$, as shown in Fig. \ref{fig:geometry}. In principle, this Green's function $G(\br,\br')$ may be obtained by an expansion in terms of the eigenfunctions of the wave operator. For that the boundary conditions need to be imposed in oblate spheroidal coordinate system, where a plate with a hole is a surface of just one coordinate being constant, see e.g. \cite{Feshbach}, but then the eigenfunctions of the wave operator are oblate spheroidal wave-functions whose handling, both analytically and numerically, is highly complicated \cite{Flammer,Chinese}. However, the required Green's function may be obtained by a far simpler method, namely, using a coordinate transformation discovered by Lord Kelvin \cite{Kelvin}, the so-called Kelvin inversion. The Kelvin inversion is a non-linear coordinate transformation, a reflection of space in a sphere of radius $S$ and centred at $\mathbf{s}$, and is defined as follows:
\begin{equation}
\boldsymbol{\mathcal{T}}[\br]=
\dfrac{S^2}{|\br-\mathbf{s}|^2}(\br-\mathbf{s})+\mathbf{s}.\label{eqn:Inversion}
\end{equation}
The geometrical properties of this transformation are most clearly presented in \cite{Jeans}. The Green's function of the Poisson equation $G(\br,\br')$ is effectively a function of two variables: the observation point $\br$ and the point where the source is located $\br'$. To apply the transformation (\ref{eqn:Inversion}) to the Green's function $G(\br,\br')$ we apply it to both of its arguments. The transformation can be used to generate new solutions of the Poisson equation from the known ones. This is possible because we have \cite{Footnote}
\begin{equation}
-\nabla^2\left[\frac{S^2}{|\br-\mathbf{s}||\br'-\mathbf{s}|}G\left(\boldsymbol{\mathcal{T}}[\br],\boldsymbol{\mathcal{T}}[\br']\right)\right]=\delta^{(3)}(\br-\br'),\label{eqn:NewPoisson}
\end{equation}
that is, the transformed Green's function $G\left(\boldsymbol{\mathcal{T}}[\br],\boldsymbol{\mathcal{T}}[\br']\right)$ augmented by an appropriate pre-factor also satisfies the Poisson equation \cite{Wermer}. Thus, the transformation
\begin{equation}
G(\br,\br')\Rightarrow \frac{S^2}{|\br-\mathbf{s}||\br'-\mathbf{s}|} G\left(\boldsymbol{\mathcal{T}}[\br],\boldsymbol{\mathcal{T}}[\br']\right)\equiv \bar{G}(\br,\br')\label{eqn:FullInversion}
\end{equation}
generates a new Green's function of the Poisson equation $\bar{G}(\br,\br')$ from an already known solution $G(\br,\br')$. The crucial point here is that, if the Green's function $G(\br,\br')$ vanishes on some surface $\sigma $, then the transformed Green's function $\bar{G}(\br,\br')$ vanishes on the transformed surface $\bar{\sigma}$, so that, by adjusting the position and the radius of the inversion sphere, one is able to obtain Green's functions for perfect reflectors of various shapes \cite{Keijo, Keijo2}. Here we follow this idea to generate the Green's function for a plate with a hole from the known case of the semi-infinite half-plane. 
\begin{figure}[t]
\includegraphics[width=8.8 cm, height=3.5 cm]{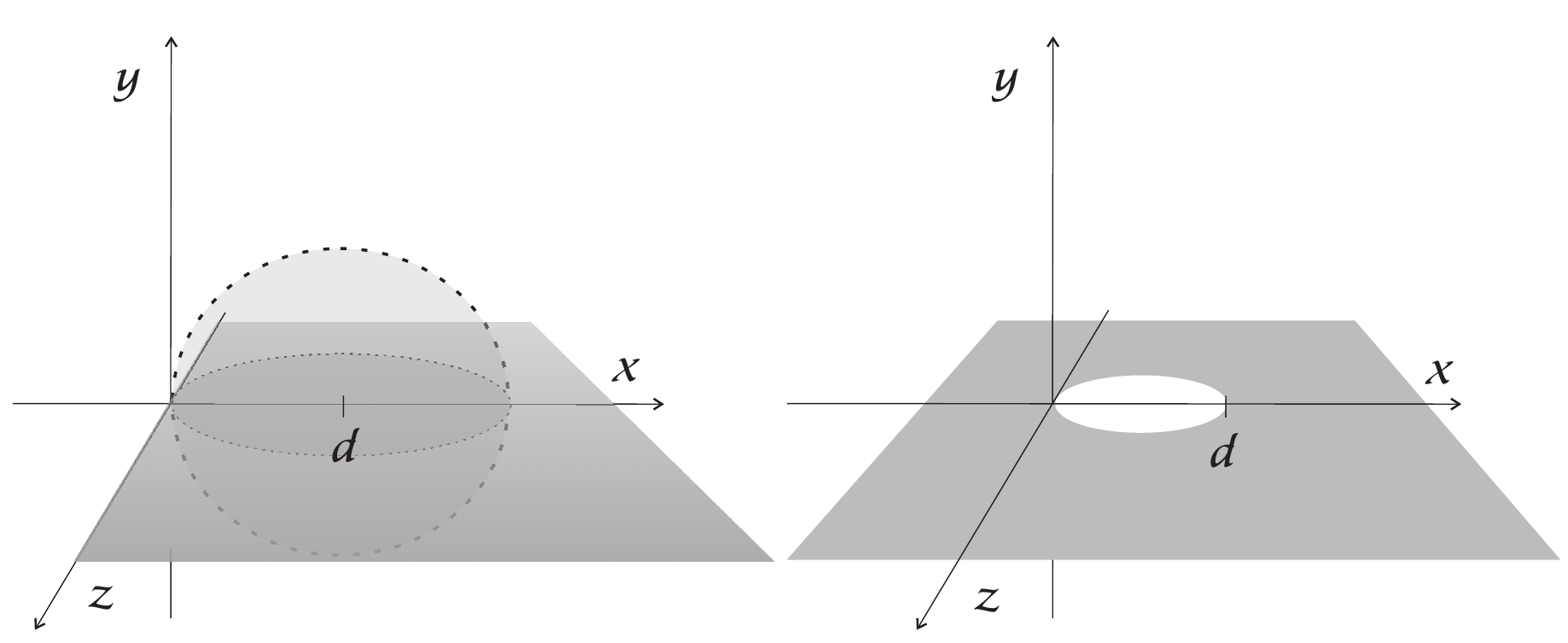}
\caption{\label{fig:Inversion} (Color online) Green's function (\ref{eqn:HalfPlane}) corresponds to a half-plane geometry when the perfect reflector occupies the $y=0\;\cap\;x\ge 0$ plane (left). Application of the Kelvin inversion (\ref{eqn:FullInversion}) with $\mathbf{s}=(d,0,0)$ and $S=d$ yields the Green's function for a plate with a hole (right).}
\end{figure}
The Green's function for a perfectly-reflecting half-plane expressed in cylindrical coordinates has been calculated in \cite{Micro1}:
\begin{eqnarray}\
G(\br,\br')&=&\frac{1}{8\pi}\left\{\frac{1}{D_-}\left[1+\frac{2}{\pi}\arctan\left(\frac{F_-}{D_-}\right)\right]\right.\nonumber\\
&-&\left.
\frac{1}{D_+}\left[1+\epsilon\frac{2}{\pi}\arctan\left(\frac{F_+}{D_+}\right)\right]
\right\}.\label{eqn:HalfPlane}
\end{eqnarray}
The formula is valid for $\br$ and $\br'$ lying on the same side of the half-plane. The abbreviations in Eq. (\ref{eqn:HalfPlane}) are:
\begin{eqnarray}
D_\mp &=& \sqrt{\rho^2+\rho'^2-2\rho\rho'\cos(\phi\mp\phi')+(z-z')^2},\nonumber\\
F_\mp &=& \sqrt{2\rho\rho'[1+\cos(\phi\mp\phi')]},\;\;\; \epsilon = {\rm sgn}[\sin(\phi+\phi')].\nonumber
\end{eqnarray}
The Green's function (\ref{eqn:HalfPlane}) satisfies equation (\ref{eqn:Poisson}) and vanishes on the surface $\phi=0$ which corresponds to the surface $y=0\;\cap\; x\ge 0$ in Cartesian coordinates, as illustrated in Fig. \ref{fig:Inversion} (left). In order to obtain the Green's function for a plate with a circular hole we apply the transformation (\ref{eqn:FullInversion}) with the centre of inversion located at the point $\mathbf{s}=(d,0,0)$ and with its radius being $S=d$. For this particular choice of $\mathbf{s}$ and $S$, the transformation maps the half-plane $\sigma=\{\br \in \mathbb{R}^3 : y=0\;\cap\;x\ge 0\}$ to a sheet with a circular hole: $\bar{\sigma}=\{\br\in \mathbb{ R}^3:y=0\;\cap\; (x-d/2)^2+z^2 \ge (d/2)^2\}$, cf. Fig. \ref{fig:Inversion} (right). Naturally we wish the coordinate system to be placed at the centre of the aperture in such a way that the $z$-axis is perpendicular to the sheet, as shown in Fig. \ref{fig:geometry}. Thus we shift and rotate the axes according to $x\rightarrow  d/2 +y,\; y\rightarrow z,\; z\rightarrow x$. It is straightforward to check that the procedure described above yields the transformed Green's function in the form as Eq. (\ref{eqn:HalfPlane}) but with
\begin{eqnarray}
F_\mp &=&\frac{\sqrt{2}}{d}\left\{\left(\rho^2+z^2-\frac{d^2}{4}\right)\left(\rho'^2+z'^2-\frac{d^2}{4}\right)\right.\nonumber\\
& \pm &
d^2zz'+
\sqrt{
\bigg[z^2+\left(\rho-\frac{d}{2}\right)^2\bigg]
\bigg[z^2+\left(\rho+\frac{d}{2}\right)^2\bigg]
}
\nonumber
\end{eqnarray}
\begin{eqnarray}
&\times &\left.
\sqrt{
\bigg[z'^2+\left(\rho'-\frac{d}{2}\right)^2\bigg]
\bigg[z'^2+\left(\rho'+\frac{d}{2}\right)^2\bigg]
}
\right\}^{1/2},\nonumber\\
D_\mp &=& \sqrt{\rho^2+\rho'^2-2\rho\rho'\cos(\phi-\phi')+(z\mp z')^2},\nonumber
\end{eqnarray}
and $\epsilon={\rm sgn}[z(\rho'^2+z'^2-d^2/4)+z'(\rho^2+z^2-d^2/4)]$. 
The resulting Green's function is valid for $\br$ and $\br'$ lying on the same side of the sheet. It satisfies Poisson's equation (\ref{eqn:Poisson}) and vanishes on the surface of the conductor, i.e. for all $\br$ with $z=0\;\cap \; x^2+y^2\ge d^2/4$. To see this it is instructive to go into spheroidal coordinates \cite{Feshbach} where for $\xi$, $\xi'$, $\eta$, $\eta'$ all positive we get
\begin{eqnarray}
F_\mp &=& \frac{d}{2}\left|\eta\eta'\pm\xi\xi' \right|\;,\ \  \epsilon =\mbox{sgn}(\xi\xi'-\eta\eta')\;,\nonumber\\
D_\mp &=& \frac{d}{2}\bigg[(\xi-\xi')^2-(\eta\mp\eta')^2+2(1+\xi\xi')(1\mp\eta\eta')\nonumber\\
& -&2\sqrt{(\xi^2+1)(1-\eta^2)(\xi'^2+1)(1-\eta'^2)} \cos(\phi-\phi')\bigg]^{1/2}.\nonumber
\end{eqnarray}
The knowledge of the Green's function now puts us into the position to calculate the atomic energy shift (\ref{eqn:EnergyShift}), 
\begin{equation}
\Delta E = -\frac{1}{16\pi^2\epsilon_0}\left(
\Xi_\rho\langle\mu^2_\rho\rangle
+\Xi_\phi\langle\mu^2_\phi\rangle
+\Xi_z\langle\mu^2_z\rangle
\right)\label{eqn:TotalShift}
\end{equation}
with
\begin{eqnarray}
\Xi_\rho &=& \frac{d\rho^2}{R_+^5 R_-^5}\left[\left(\rho^2+z^2-\frac{d^2}{4}\right)^2-d^2z^2\right]\nonumber\\
&+&\frac{d^3}{6}\frac{1}{R_+^3 R_-^3}+\frac{1}{4z^3}\bigg[\frac{\pi}{2}+\arctan\left(\dfrac{\rho^2+z^2-d^2/4}{dz}\right)\nonumber\\
&+&\frac{dz}{R_+^4 R_-^4}\left(\rho^2-z^2-\frac{d^2}{4}\right)^2\left(\rho^2+z^2-\frac{d^2}{4}\right)\bigg],\label{eqn:ShiftRho}\\
\Xi_\phi &=& \frac{d^3}{6}\frac{1}{R_+^3 R_-^3}+\frac{1}{4z^3}\bigg[\frac{\pi}{2}+\arctan\left(\dfrac{\rho^2+z^2-d^2/4}{dz}\right)\nonumber\\
&\;&\hspace{2 cm}+\frac{dz}{R_+^2 R_-^2}\left(\rho^2+z^2-\frac{d^2}{4}\right)\bigg],\label{eqn:ShiftPhi}\\
\Xi_z &=& \frac{d}{R_+^5 R_-^5}\bigg[z^2\left(\rho^2+z^2+\frac{d^2}{4}\right)^2
-\frac{d^2}{4}\left(\rho^2-z^2-\frac{d^2}{4}\right)^2\bigg]\nonumber\\
&+&\frac{d^3}{6}\frac{1}{R_+^3 R_-^3}+\frac{1}{2z^3}\bigg[\frac{\pi}{2}+\arctan\left(\dfrac{\rho^2+z^2-d^2/4}{dz}\right)\nonumber\\
&+&\frac{dz}{R_+^2 R_-^2}\left(\rho^2-z^2-\frac{d^2}{4}\right)
+\frac{2d\rho^2 z^3}{R_+^4 R_-^4}\left(\rho^2+z^2-\frac{d^2}{4}\right)
\bigg],\nonumber\\\label{eqn:ShiftZ}
\end{eqnarray}
where we have defined
\begin{equation}
R_{\pm}=\left[\left(\rho\pm\frac{d}{2}\right)^2+z^2\right]^{1/2}.
\end{equation}
The result (\ref{eqn:TotalShift}) together with Eqs. (\ref{eqn:ShiftRho})-(\ref{eqn:ShiftZ}) is, as far as electrostatics is concerned, exact. It applies to atoms whose distance from the surface is much smaller than the wavelength of the dominant dipole transition.
\begin{figure}[t]
\includegraphics[width=6.7 cm, height=6.7 cm]{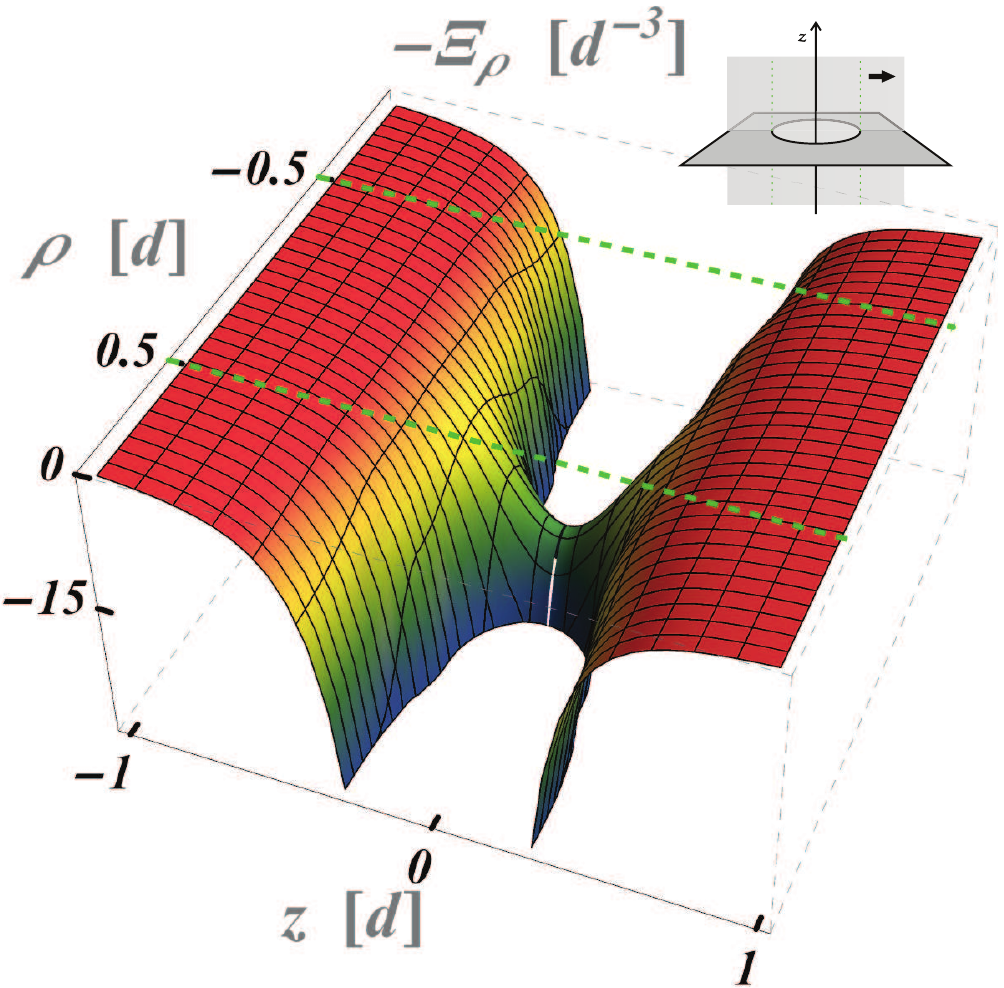}
\caption{\label{fig:PotRho} (Color online) Interaction potential between a plate with a hole of diameter $d$ and a dipole polarized in the $\rho$-direction, cf. Fig \ref{fig:geometry}. The inset illustrates the plane on which $\Xi_\rho$ is evaluated and the orientation of the dipole.}
\end{figure}
\begin{figure}[t]
\includegraphics[width=6.7 cm, height=6.7 cm]{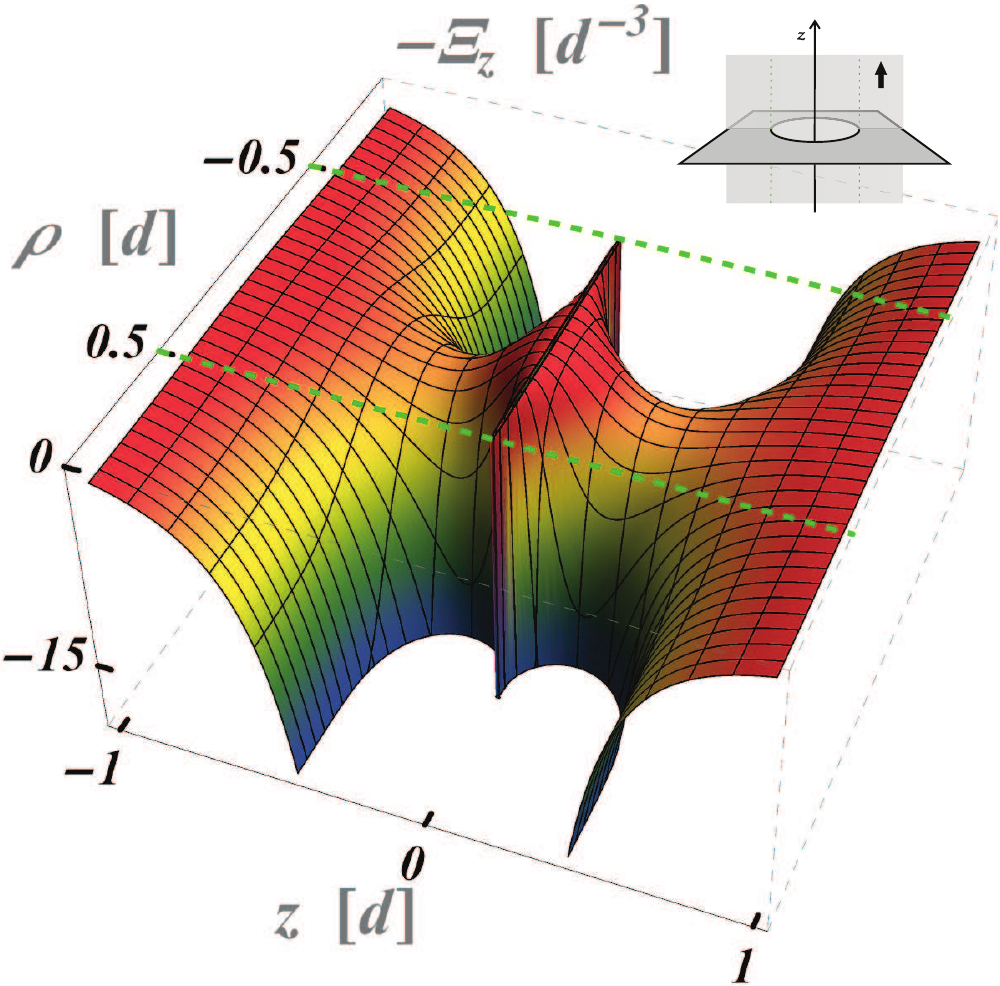}
\caption{\label{fig:PotZ} (Color online) Interaction potential between a plate with a hole of diameter $d$ and a dipole polarized in the $z$-direction, cf. Fig \ref{fig:geometry}. The inset illustrates the plane on which $\Xi_z$ is evaluated and the orientation of the dipole.}
\end{figure}
\begin{figure}[t]
{\includegraphics[width=7.0 cm, height=4 cm]{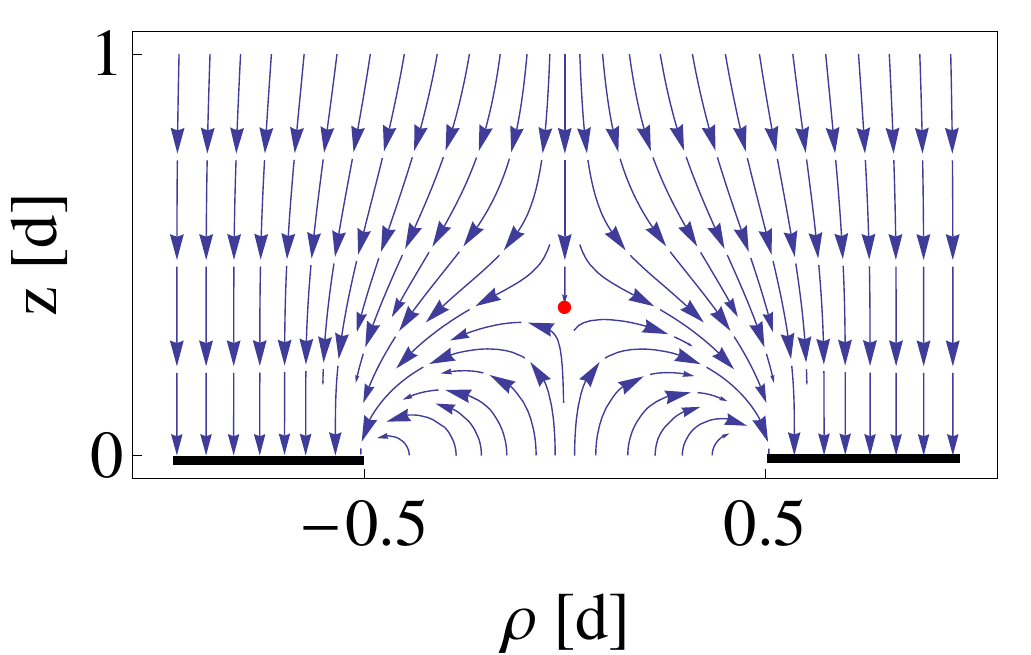}
\includegraphics[width=7.0 cm, height=4 cm]{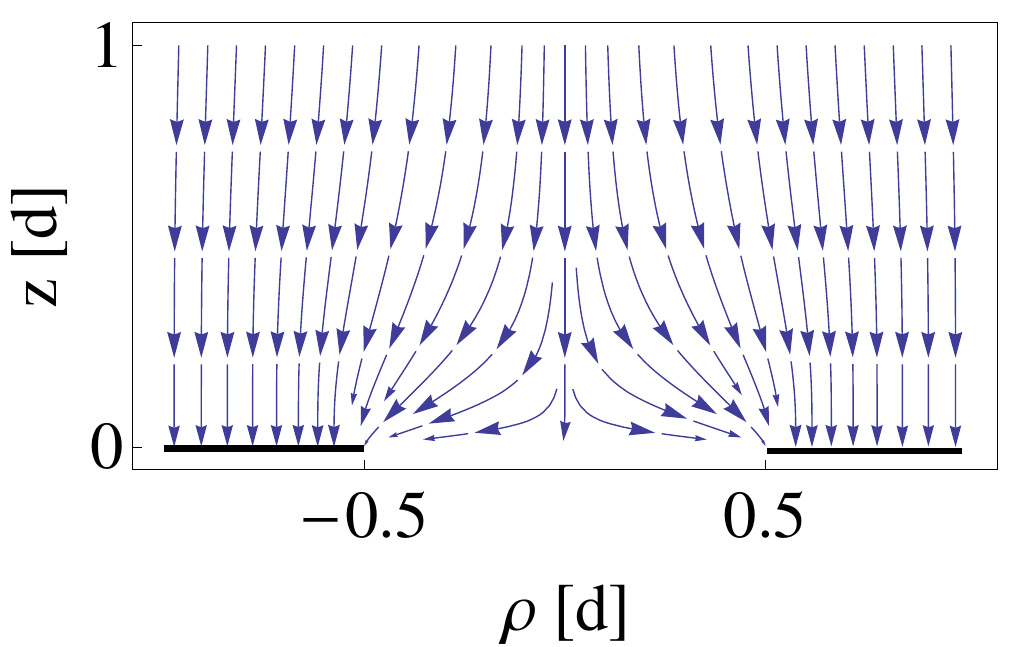}}
\caption{\label{fig:Force} (Color online) The direction of the Casimir-Polder force acting on a $z$-polarized atom interacting with a plate with a hole of diameter $d$ (top). The (red) dot shows the location of the saddle point at $z\approx 0.3711 d$. The direction of the Casimir-Polder force acting on isotropically polarized atom does not feature unusual patterns (bottom).}
\end{figure}
The simplicity of Eqs. (\ref{eqn:ShiftRho})-(\ref{eqn:ShiftZ}) allows us to visualize the Casimir-Polder potential felt by an atom. As anticipated, for the $z$-polarized atomic dipole, the shift vanishes in the plane of the sheet, cf. Fig. \ref{fig:PotZ}, which can be confirmed also analytically. The potential evaluated along lines of constant $\rho<d/2$ indeed has a minimum above and below the aperture. Interestingly, there is in principle a region where a $z$-polarized dipole is repelled from the microstructure, as has been concluded on the basis of numerical calculations along the line $\rho=0$ \cite{Levin}. One of the benefits of our exact analytical result for the complete $\rho$-$z$ plane is to show that motion along the line $\rho=0$ is unstable and upon the slightest perturbation the dipole is attracted towards the edge of the aperture, cf. Fig. \ref{fig:Force}. Thus, it is more appropriate to speak of attraction along unusual paths rather than of repulsion. The location of the saddle points and the magnitude of the shifts there can be estimated to be
\begin{equation}
z_{\rm s}\approx \pm 0.3711 d,\;\;\;\Xi_z(\rho=0,z=|z_{\rm s}|,d)\approx\frac{4.0622}{d^3}.\label{eqn:SaddleZ}
\end{equation}
The potential experienced by the atom polarized in the $\rho$ direction does not display the same peculiarities as that of a $z$-polarized one. We plot it in Fig. \ref{fig:PotRho} and stress that the potential for the $\phi$-polarized atom looks very similar. In fact, both functions, $\Xi_\rho$ and $\Xi_\phi$, coincide at $\rho=0$ and feature a saddle point only at the centre of the aperture where we have
\begin{equation}
\Xi_{\rho,\phi}(\rho=0,z=0,d)\approx\frac{21.3333}{d^3}.\label{eqn:SaddleRho}
\end{equation}
Comparing Fig. \ref{fig:PotRho} with Fig. \ref{fig:PotZ}, together with Eqs. (\ref{eqn:SaddleZ})-(\ref{eqn:SaddleRho}), we note that although the potential strongly depends on the polarization of the dipole the characteristic features of $\Xi_{\rho,\phi}$ are by far more pronounced than those of $\Xi_z$. Thus, for ground-state atoms with isotropic polarizabilities the Casimir-Polder force will not display any unusual behaviour, cf. Fig. \ref{fig:Force}. However, for cold polar molecules that by some other means are being confined to move along the $z$ axis, one could in principle use this system for population-sensitive trapping: molecules with dipoles along $z$ would trap above and below the plane of the aperture, whereas molecules with dipoles parallel to this plane would trap right at the centre of the aperture. The only drawback to bear in mind is that the relative depth of the minima of the two potentials is fixed to about 5.25 independent of the size of the aperture, so that the population of molecules trapped with dipoles along $z$ would be smaller. 

It is worth noting that a similar polarization dependence of the dipole-surface interaction would be expected if one were to consider a sheet of a conductor with a long slit rather than a circular aperture. Such a geometry might lend itself to polarization-sensitive guiding and scattering of cold polar molecules.

The results presented here do not rely on the optical properties of the surfaces but rather on their geometry. Thus the perfect-reflector model considered here is fully appropriate. In practice the most important shortcoming of the model is the assumption of zero thickness of the conductor, which is the reason for the $z$-polarized dipole not to couple to sheet when it is located within the aperture. For cold atoms this condition is difficult to satisfy in practice, even in the era of graphene, and in reality we should expect the thickness of the material to come into play and alter the interaction. Numerical calculations have given estimates of how much the potential wells are being reduced by finite thickness of the material \cite{Levin}.
\begin{acknowledgments}
It is a pleasure to acknowledge discussions with Mark Fromhold.
\end{acknowledgments}


\vspace{-.8 cm}
\begin{thebibliography}{99}
\bibitem{CP}{H.B.G.~Casimir, D.~Polder, Phys. Rev. 73, {\bf 360} (1948).}
\bibitem{Micro1}{C.~Eberlein, R.~Zietal, Phys. Rev. A {\bf 75}, 032516(2007).}
\bibitem{Micro2}{C.~Eberlein, R.~Zietal, Phys. Rev. A {\bf 80}, 012504(2009).}
\bibitem{Levin}{M.~Levin, A.P.~McCauley, A.W.~Rodriguez, M.T.~Homer Reid, S.G.~Johnson, Phys. Rev. Lett. {\bf 105}, 090403 (2010).}
\bibitem{Feshbach}{P.~M.~Morse, H.~Feshbach, \textit{Methods of Theoretical Physics, Vol. 2}, (McGraw-Hill Book Company, Inc., New York, 1953).}
\bibitem{Flammer}{C.~Flammer, \textit{Spheroidal Wave Functions}, (Stanford University Press, Stanford, California, 1957.)}
\bibitem{Chinese}{L.~Li, M.~Leong, T.~Yeo, P.~Kooi, K.~Tan, Phys. Rev. E, {\bf 58}, 6792(1998).}
\bibitem{Kelvin}{W.~Thomson (Lord Kelvin), P.~G.~Tait, Journal de Mathematiques Pures et Appliqu\'es, {\bf 12}, 256 (1847).}
\bibitem{Jeans}{J.~H.~Jeans, \textit{Mathematical Theory of Electricity and Magnetism}, Ch. VIII,  (Cambridge University Press, 1908).}
\bibitem{Footnote}{The RHS of Eq. (\ref{eqn:NewPoisson}) contains an additional term proportional to $\delta^{(3)}(\br-\mathbf{s})\;G\left(\boldsymbol{\mathcal{T}}[\br],\boldsymbol{\mathcal{T}}[\br']\right)$, which vanishes provided $G\left(\boldsymbol{\mathcal{T}}[\br],\boldsymbol{\mathcal{T}}[\br']\right)$ vanishes for $\boldsymbol{\mathcal{T}}[\br]\rightarrow\infty$.} 
\bibitem{Wermer}{J.~Wermer, \textit{Potential Theory}, Lecture Notes in Mathematics 408, (Springer-Verlag, New York, 1974).}
\bibitem{Keijo}{Proceedings of the XXVIIth URSI General Assembly in Maastricht, Commission B (Fields and Waves)-B1 Electromagnetic Theory paper B1.P.8 (741) (August 2002).}
\bibitem{Keijo2}{K.~Nikoskinen, H.~Wall\'en, IEE Proc.-Sci. Meas. Technol., {\bf 153}, 174 (2006).}
\end{thebibliography}
\end{document}